
\input harvmac.tex
\Title{CTP/TAMU-81/92}{{A Comment on the Stability of String Monopoles}
\footnote{$^\dagger$}{Work supported in part by NSF grant PHY-9106593.}}

\centerline{
Ramzi~R.~ Khuri\footnote{$^*$}{Supported by a World Laboratory Fellowship.}}
\bigskip\centerline{Center for Theoretical Physics}
\centerline{Texas A\&M University}\centerline{College Station, TX 77843}

\vskip .3in
In recent work a multimonopole solution of heterotic string theory was
obtained. The monopoles are noted to be stable, in contrast with analogous
solutions of Einstein-Maxwell or Yang-Mills-dilaton theory. The existence of
this and other classes of stable solitonic solutions in string theory thus
provides a possible test for low-energy string theory as distinct
from other gauge + gravity theories.

\Date{12/92}
\lref\quartet{D.~J.~Gross,
J.~A.~Harvey, E.~J.~Martinec and R.~Rohm, Nucl. Phys. {\bf B256} (1985) 253.}

\lref\dine{M.~Dine, Lectures delivered at
TASI 1988, Brown University (1988) 653.}

\lref\chad{J.~M.~Charap and M.~J.~Duff, Phys. Lett. {\bf B69} (1977) 445.}

\lref\brone{E.~A.~Bergshoeff and M.~de Roo, Nucl. Phys. {\bf B328} (1989) 439.}

\lref\brtwo{E.~A.~Bergshoeff and M.~de Roo, Phys. Lett. {\bf B218} (1989) 210.}

\lref\ccrk{C.~G.~Callan and R.~R.~Khuri, Phys. Lett. {\bf B261} (1991) 363.}

\lref\mono{R.~R.~Khuri, Phys. Lett. {\bf B294} (1992) 325.}

\lref\monscat{R.~R.~Khuri, Phys. Lett. {\bf B294} (1992) 331.}

\lref\monex{R.~R.~Khuri, Nucl. Phys. {\bf B387} (1992) 315.}

\lref\nair{K.~Lee, V.~P.~Nair and E.~J.~Weinberg, Phys. Rev. Lett. {\bf 68}
(1992) 1100.}

\lref\geo{R.~R.~Khuri, ``Geodesic Scattering of Solitonic Strings",
Texas A\&M preprint CTP/TAMU-79/92.}

\lref\scat{R.~R.~Khuri, ``Classical Scattering of Macroscopic Strings",
Texas A\&M preprint CTP/TAMU-80/92.}

\lref\dfluone{M.~J.~Duff and J.~X.~Lu, Nucl. Phys. {\bf B354} (1991) 141.}

\lref\dflutwo{M.~J.~Duff and J.~X.~Lu, Nucl. Phys. {\bf B354} (1991) 129.}

\lref\bogo{E.~B.~Bogomolnyi, Sov. J. Nucl. Phys. {\bf 24} (1976) 449.}

\lref\chsone{C.~G.~Callan, J.~A.~Harvey and A.~Strominger, Nucl. Phys.
{\bf B359} (1991) 611.}

\lref\chstwo{C.~G.~Callan, J.~A.~Harvey and A.~Strominger, Nucl. Phys.
{\bf B367} (1991) 60.}

\lref\strom{A.~Strominger, Nucl. Phys. {\bf B343} (1990) 167.}

\lref\felce{A.~G.~Felce and T.~M.~Samols, ``Low-Energy Dynamics of String
Solitons" NSF-ITP-92-155, November 1992.}

\lref\jjj{J.~P.~Gauntlett, J.~A.~Harvey and J.~T.~Liu, ``Magnetic
Monopoles in String Theory" IFP-434-UNC, November 1992.}

\lref\bizon{P.~Bizon, ``Saddle Point Solutions in Yang-Mills-Dilaton
theory" UWThPh-1992-46, September 1992.}

\lref\lavrel{G.~Lavrelashvili and D.~Maison, ``Static Spherically Symmetric
Solutions of a Yang-Mills Field Coupled to a Dilaton", MPI-Ph/92-62,
August 1992.}

\lref\bart{R.~Bartnik and J.~McKinnon, Phys. Rev. Lett. {\bf 61} (1988)
141.}

\lref\dghrr{A.~Dabholkar, G.~Gibbons, J.~A.~Harvey and F.~Ruiz Ruiz,
Nucl. Phys. {\bf B340} (1990) 33.}

\lref\grosp{D.~J.~Gross and M.~J.~Perry, Nucl. Phys. {\bf B226} (1983) 29.}

\lref\sork{R.~D.~Sorkin, Phys. Rev. Lett {\bf 51} (1983) 87.}

\def\sqr#1#2{{\vbox{\hrule height.#2pt\hbox{\vrule width
.#2pt height#1pt \kern#1pt\vrule width.#2pt}\hrule height.#2pt}}}

\def\met {g_{\mu\nu}}

In recent work\mono, a multimonopole solution of
heterotic string theory was presented. An analogous solution in Yang-Mills
field
theory was found to have divergent action near
each source. In the string solution, however, the divergences from the
Yang-Mills sector are precisely cancelled by those from the gravity sector,
so that the action is finite and easily computed\refs{\mono,\monex}.

In this letter we comment briefly on the stability of this solution.
We note that the string monopole solution, inasmuch as it saturates a
Bogomol'nyi bound between ADM mass and charge, is stable. The stability of the
resultant low-energy YM + dilaton gravity field theory contrasts with the
recently demonstrated instability
of analogous non-string Einstein-Maxwell or YM-dilaton solutions and thus
represents a possible low-energy test for string theory.
We also briefly comment on the dynamics of these solitons.

The bosonic fields for the self-dual multimonopole solution of heterotic
string theory with zero background fermi fields are given by\mono
\eqn\anstz{\eqalign{\met&=e^{2\phi}\delta_{\mu\nu},\quad g_{ab}=\eta_{ab},\cr
H_{\mu\nu\lambda}&=\pm\epsilon_{\mu\nu\lambda\sigma}\partial^\sigma\phi,\cr
e^{2\phi}&=e^{2\phi_0}f,\cr
A_\mu&=i \overline{\Sigma}_{\mu\nu}\partial_\nu \ln f,\cr}}
where $\mu,\nu,\lambda,\sigma=1,2,3,4$, $a,b=0,5,6,7,8,9$,
$\overline{\Sigma}_{\mu\nu}=\overline{\eta}^{i\mu\nu}(\sigma^i/2)$
for $i=1,2,3$ ($\sigma^i$, $i=1,2,3$ are the $2\times 2$ Pauli matrices) where
\eqn\hfeta{\eqalign{\overline{\eta}^{i\mu\nu}=-\overline{\eta}^{i\nu\mu}
&=\epsilon^{i\mu\nu},\qquad\qquad \mu,\nu=1,2,3,\cr
&=-\delta^{i\mu},\qquad\qquad \nu=4 \cr}}
and where
\eqn\fdmono{f=1+\sum_{n=1}^N{m_n\over |\vec x - \vec a_n|},}
where $m_n$ is the charge and $\vec a_n$ the location in
the three-space $(123)$ of the $n$th monopole. The anti-self-dual solution is
similar, with the $\delta$-term in \hfeta\ changing sign. This solution
was shown to have multimonopole structure\mono\ in the four-space $(0123)$,
each source having topological charge $Q=1$ and magnetic charge $m=1/g$,
where $g$ is the YM coupling constant. The four-dimensional metric line-element
strongly resembles that of the Kaluza-Klein monopole \refs{\grosp,\sork}. We
argued in \mono\ that this solution is exact to all orders in $\alpha'$.

If we make the identification $\Phi\equiv A_4$, then the gauge and Higgs
fields may be simply written in terms of the dilaton as
\eqn\stmono{\eqalign{\Phi^a&=-{2\over g}\delta^{ia}\partial_i\phi,\cr
A_k^a&=-{2\over g}\epsilon^{akj}\partial_j\phi\cr}}
for the self-dual solution. For the anti-self-dual solution, the Higgs
field simply changes sign. A toroidal compactification
can be adopted so that we consider the dynamics of our solution
in the spacetime $(0123)$. As usual, the existence of a
static multi-soliton solution depends on the ``zero force'' condition.

A similar solution for Yang-Mills-dilaton theory was found in \bizon,
with analogous dilaton behaviour and corresponding to a Dirac magnetic
monopole with unit magnetic charge. It was noted that this solution has
infinitely many unstable modes, but has finite action resulting from
the cancellation between divergences stemming from the YM field and
dilaton respectively. In fact, magnetically charged solutions to the
coupled Einstein-Maxwell equations are typically found to be unstable
\refs{\bart,\nair,\lavrel}, with or without the presence of a dilaton.

The ansatz of \stmono\ represents an infinite-action solution
of YM + scalar field theory in $3+1$ dimensions which satisfies
the Bogomol'nyi bound $G_{ij}^a=\epsilon_{ijk}D_k\Phi^a$. Since the action
is finite away from the singularity, the Bogomol'nyi bound guarentees
stability of the solution outside arbitrary finite enclosures around each
source. Since the string solution has finite action owing to the
cancellation between gauge and gravitational divergences,
we can use a Bogomol'nyi bound to demonstrate stability everywhere.
We see this by noting that the ADM mass of the multimonopole solution
written in terms of the canonical metric saturates a Bogomol'nyi bound
in terms of the charges of the string monopoles\jjj
\eqn\stmonbog{M={2\pi\over \kappa^2}e^{\phi_0/2}\sum_{i=1}^N m_i.}
Thus the string multimonopole solution sits at a minimum energy point and is
automatically stable against perturbations independent of $x_4$ (presumably
if we allow perturbations depending on the compactified direction the
string monopoles will decay into heterotic instanton fivebranes
\refs{\dfluone\dflutwo\strom\chsone{--}\chstwo}, which possess a higher
symmetry\foot{This was pointed out to me by Jianxin Lu.}).
As an exercise, one can explicitly show that there are no unstable modes
of the string monopole away from the singularity, from either the gauge or
gravitational sectors. The same divergence cancellation feature which
leads to a finite action solution then allows us to circumvent the
singularity in each sector when considering the stability of the string
solution.

Since the divergence cancellation in the action is also possessed by the
unstable solution in \bizon, however, this property is clearly insufficient in
itself to explain the stability of the string solution, which is
rooted in the (4,4) superconformal invariance of its underlying sigma-model
\refs{\chsone,\chstwo}. It is also linked to the existence of a zero-dynamical
force condition obeyed by this and other classes of heterotic multi-soliton
solutions \refs{\dghrr\ccrk\felce\geo{--}\scat}, in addition to the usual
zero-static force condition. Since the string solutions all saturate
Bogomol'nyi bounds and are hence stable, their existence in suitably
compactified form provides a possible test for string theory as distinct
from other gauge + gravity models.

A study of the low-energy dynamics of the string monopoles was done in
\monscat,
where it was found from both a test-monopole approach and a computation of
the Manton metric on moduli space that the monopoles scatter trivially in
the low-energy limit. The latter method relied on the construction of an
$O(\beta)$ solution to the constraint equations of motion, which had the
piecewise Lorentz boosted form
\eqn\orderbeta{\eqalign{e^{2\phi(\vec x,t)}&=1+\sum_{n=1}^N{m_n\over
|\vec x - \vec a_n(t)|},\cr g_{00}&=-1,\qquad g^{00}=-1,\qquad
g_{ij}=e^{2\phi}\delta_{ij},\qquad g^{ij}=e^{-2\phi}\delta_{ij},\cr
g_{0i}&=-\sum_{n=1}^N{m_n\vec\beta_n\cdot \hat x_i\over |\vec x - \vec
a_n(t)|},\qquad g^{0i}=e^{-2\phi}g_{0i},\cr
H_{ijk}&=\epsilon_{ijkm}\partial_m e^{2\phi},\cr
H_{0ij}&=\epsilon_{ijkm}\partial_m g_{0k}=\epsilon_{ijkm}\partial_k
\sum_{n=1}^N{m_n\vec\beta_n\cdot \hat x_m\over |\vec x - \vec a_n(t)|},\cr}}
where $i,j,k,m=1,2,3,4$ and we use a
flat space $\epsilon$-tensor. Note that $g_{00}$, $g_{ij}$ and $H_{ijk}$ are
unaffected to order $\beta$. According to the conjecture of Bergshoeff and
de Roo\refs{\brone,\brtwo,\mono}, from the exactness condition
$A_\mu=\Omega_{\pm\mu}$\refs{\chad,\dine} (where
$\Omega_{\pm\mu}$ is the generalized connection defined in
\mono), the higher order in $\alpha'$ terms drop out from the action.
The contributions of these terms
also cancel in the equations of motion, provided we assume a similar
$O(\beta)$ solution for the YM field $A_\mu$ to balance the piecewise-boosted
generalized connection $\Omega_{\pm\mu}$. We therefore argue that our
low-energy scattering analysis in \monscat\ is also exact, suggesting
that the dynamic YM force for these string monopoles is precisely cancelled by
the dynamic gravity sector force to all orders in $\alpha'$. This argument
can equally well be applied to the heterotic instanton fivebranes.

\bigbreak\bigskip\bigskip\centerline{{\bf Acknowledgements}}\nobreak
I would like to thank M.~Duff, A.~Felce, Z.~Khviengia, J.~Lu, V.~P.~Nair,
M.~Ortiz and E.~Weinberg for helpful discussions.

\vfil\eject
\listrefs
\bye